\documentclass[RNAAS]{aastex631} 

\usepackage{amsmath}

\usepackage{mathtools}

\newcommand\myeq{\stackrel{\mathclap{\normalfont\mbox{!}}}{=}}

\begin{document}

\title{Detecting the Undetected: Overcoming Biases in Gravitational-Wave Population Studies}

\author[0000-0003-4083-6390]{Ryan Raikman}
\affiliation{Carnegie Mellon University, 5000 Forbes Ave, Pittsburgh, PA, 15213, USA}
\affiliation{Département d'Astronomie, Université de Genève, Chemin Pegasi 51, CH-1290 Versoix, Switzerland}

\author[0000-0002-3439-0321]{Simone Bavera}
\affiliation{Département d'Astronomie, Université de Genève, Chemin Pegasi 51, CH-1290 Versoix, Switzerland}
\affiliation{Gravitational Wave Science Center (GWSC), Université de Genève, 24 quai E. Ansermet, CH-1211 Geneva, Switzerland}

\author{Tassos Fragos}
\affiliation{Département d'Astronomie, Université de Genève, Chemin Pegasi 51, CH-1290 Versoix, Switzerland}
\affiliation{Gravitational Wave Science Center (GWSC), Université de Genève, 24 quai E. Ansermet, CH-1211 Geneva, Switzerland}

\begin{abstract}

In the flourishing field of gravitational-wave astronomy, accurately inferring binary black hole merger formation channels is paramount. The Bayesian hierarchical model selection analysis offers a promising methodology \citep[see, e.g., \textit{``One Channel to Rule Them All''},][]{2021ApJ...910..152Z}. However, recently, \citet{2023arXiv230703129Q} highlighted a critical caveat: observed channels absent in known models can bias branching fraction estimates. In this research note, we introduce a test to detect missing channels in such analyses. Our findings show a commendable success rate in identifying these elusive channels. Yet, in scenarios where missing channels closely overlap with recognized ones, discerning the difference remains challenging.

\end{abstract}

\section{Introduction} \label{sec:intro}

Gravitational-wave observations of binary black hole (BBH) mergers offer an invaluable benchmark for population synthesis studies, aiming to unravel the formation mechanisms leading to these exotic objects. By utilizing binary population tools like \texttt{POSYDON} \citep{2023ApJS..264...45F}, researchers can predict the observable properties of merging BBHs. Different formation scenarios imprint different signatures in the intrinsic distributions of the observable BBH properties such as their chirp mass, mass ratio, effective spin parameter, and merger redshift. 

The discovery of merging compact binaries with GWs has led to a number of studies that attempt to infer the branching fractions of relevant formation channels \citep[e.g.,][]{2015ApJ...810...58S,Zevin_2017,2018ApJ...854L...9F}. More recently, \citet{2021ApJ...910..152Z} expanded this methodology to include five formation channels modeled self-consistently.
Nonetheless, the recent study of \citet{2023arXiv230703129Q} cautions about substantial biases in branching fraction computations if an unaccounted-for formation channel exists in observed data.

This research note introduces a test to detect the presence of an unknown, unaccounted-for, formation channel in the Bayesian hierarchical model selection analysis. In the next section, we showcase the test procedure using as a reference the five formation channels used in the analysis of \citet{2021ApJ...910..152Z} and \citet{2023arXiv230703129Q}.

\section{Procedure and Results} \label{sec:procedure}

Our objective is to retrieve branching fractions from observed merging BBH populations. This entails mapping the observed distribution within a set of known population models. In a non-orthonormal basis scenario, expected coefficients of the branching fractions are determined via a system of linear equations. We denote the observed population, $P_O$, as a normalized probability density function (PDF) composed of formation channels $p_i$ for $i=1...N$, with corresponding branching fractions $b_i$, i.e. $P_O = \sum_{i=1}^N b_i p_i(\vec{x})$. Using the observed population and the modeled formation channels, we can compute the dot product of the observed population with a given formation channel model, $R_j$, and the dot products of the different modeled formation channels, $M_{ij}$, as defined below. Here, all PDFs are defined within an $n$-dimensional space, $\vec{x} \in \mathbb{R}^n$. We have,
\begin{equation}
    R_j \equiv \int_{ \mathbb{R}^n } P_O(\vec{x}) p_j(\vec{x}) \, d^nx \myeq  \int_{ \mathbb{R}^n } \left( \sum_{i=1}^N b_i p_i(\vec{x}) \right) p_j(\vec{x}) \, d^nx=  \sum_{i=1}^N  b_i \int_{ \mathbb{R}^n }  p_i(\vec{x}) p_j(\vec{x}) \, d^nx = \sum_{i=1}^N M_{ij} b_i
\end{equation}
where, for convenience, we define $M_{ij} \equiv  \int_{ \mathbb{R}^n }  p_i(\vec{x}) p_j(\vec{x}) \, d^nx$. The distributions $P_O$ and $p_{i}$ can be derived by fitting PDFs to the observed BBH population and the Monte Carlo-sampled formation channel models through methods like kernel density estimation (KDE), normalizing flows, etc. Once $M_{ij}$ and $R_j$ are known, one can solve the system of linear equation with respect to the branching fractions $b_i$. In the case of non-degenerate formation channels $p_i$, the expansion of $P_O$ in terms of $b_ip_i(\vec{x})$ is unique in terms of the branching fractions $b_i$. However, in cases where high degeneracy exists in the formation channels, the solutions become non-unique, and the algorithm can struggle to recognize if a degenerate formation channel is missing.

To validate our approach, we employed models for the formation of merging BBHs including different formation channels from isolated binary evolution and dynamical formation as released by \citet{2021ApJ...910..152Z}. These channels include: common envelope (CE) and stable mass transfer (SMT) channels \citep{2021A&A...647A.153B}, chemically homogeneous evolution \citep[CHE;][]{2020MNRAS.499.5941D}, nuclear star cluster \citep[NSC;][]{2019MNRAS.486.5008A}, and globular cluster \citep[GC;][]{2019PhRvD.100d3027R}. To illustrate how model uncertainties are treated, for the CE channel, we consider five different values for the efficiency of CE ejection, $\alpha_\mathrm{CE} \in [0.2,0.5,1,2,5]$, while for all models we consider four different values for the birth spin of isolated black holes as a proxy for the efficiency of angular momentum transport in the interior of the black hole progenitor stars, namely $\chi \in [0.0, 0.1, 0.2, 0.5]$. These parameters were arbitrarily taken from a larger set of model uncertainties and echo physical uncertainties in stellar and binary evolution simulations. 

For each parameter set, we conducted five simulation rounds. In each round, we generated a sample of $10^5$ BBHs as our observed population by drawing from all channels, while we omit one channel. This is done in the detectable space, with the detection effects accounted for. The genuine branching fraction $b_u$ of the excluded channel ranged between 0 and 1, and the rest bore equal branching fractions totaling $1-b_u$. The aim of our test is to detect disparities between recognized and inherent channels.
It must be noted that the analysis presented here uses a large sample size to obtain an accurate measurement of the orthogonality of distributions, which implies that the method will only be of use with the next-generation gravitational wave detectors with much greater number of detections than currently available at the writing of this research note. The need for large sample sizes comes from the difficulty of computationally fitting a PDF to the observed data. Alliteratively, improved techniques for fitting a PDF, rather than the KDE method used here, might reduce the samples required for accurate results. 
Finally, the analysis presented here, does not include measurement uncertainties in the mock observed population. A careful treatment of measurement uncertainties in illustrated test is left for future work.

\begin{figure}[h]
\centering
\includegraphics[width=1.0\textwidth]{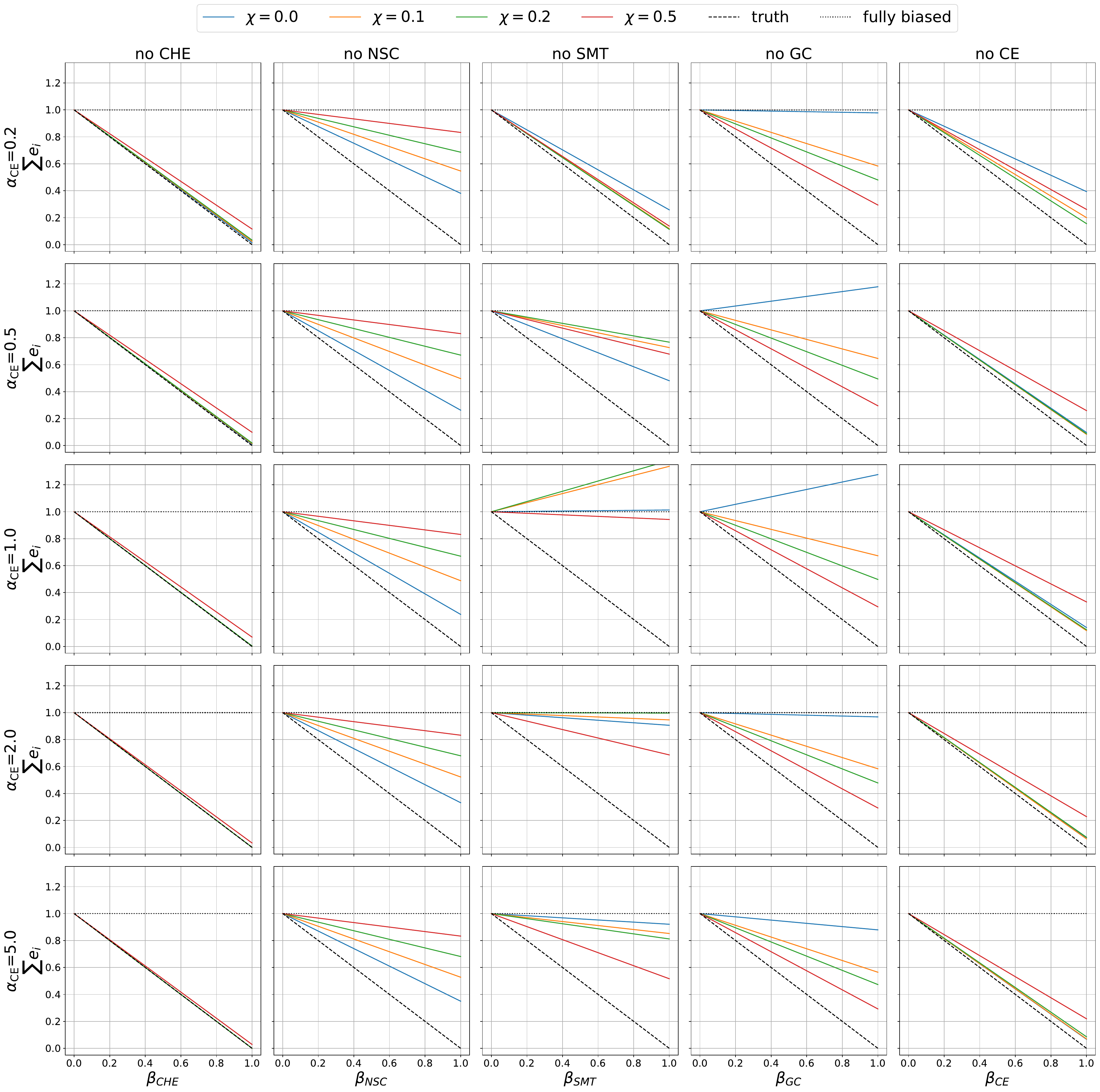}
\caption{Biasing plot - ability of our method to measure the bias caused by removing a certain formation channel from the set of known distributions. Across different values of $\alpha_\mathrm{CE}$ and $\chi$, we show the sum of estimated branching fractions from the known channels as a function of the branching fraction of the unknown channel. The optimal case is the sloped dashed black line, where the branching fraction of the unknown channel is perfectly predicted. The flat dotted black line represents no knowledge of the removed formation channels, and an overestimation of the prevalence of the known channels.} 
\label{fig:biasing_plot}
\end{figure}

We present the test results in Figure~\ref{fig:biasing_plot}. The dashed black curve serves as our baseline, denoting no bias; i.e., an unknown channel doesn't skew inferred branching fractions. This benchmark is predominantly observed in the cases of absent CE and CHE channels, implying these channels exhibit distinct signatures that can tell them apart from the others, e.g., their characteristic effective spin parameter distributions. Conversely, for other instances --- the specific scenarios without NSC or GC --- the curves fluctuate between the no-bias line ($\sum e_i=1-\beta$) and the fully biased line ($\sum e_i=1$). While this suggests that the existence of certain unseen channels can inflate assessed fractions, they never reach complete bias, suggesting our method can still detect missing channels. However, it might undervalue the branching fraction of the unknown channel or overvalue those of known ones. Exceptionally, for a few cases like in the $\alpha_\mathrm{CE}=0.2$ and $\chi=0$ scenario without GC, the method falters, barely diverging from the fully biased curve ($\sum e_i=1$). This anomaly can be attributed to degeneracies between formation channels predicting a similar observable distribution of the considered BBH observational properties: chirp mass, mass ratio, effective spin parameter and redshift of merger. Hence, the procedure presented here can be used as a test to reject the results of a Bayesian hierarchical model selection analysis but not to confirm them.

\section{Conclusion} \label{sec:conclusion}

We introduced a procedure to assess if branching fraction estimations from studies of observed gravitational-wave populations might be skewed by unaccounted formation channels. When no degeneracies exist between channels, discrepancies between the sum of branching fractions of known populations and unity emerge. This highlights potential overlooked channels. Current methodologies \citep[e.g.,][]{2021ApJ...910..152Z} assume comprehensive knowledge of formation channels, risking biases. The present research note proposes a test to detect the presence of such bias.

A thorough interpretation of branching fractions from hierarchical Bayesian model selection remains crucial, ensuring that astrophysical insights are accurate and not artifacts of a biased analysis.

\section*{Code and Data Release Statement}
All necessary code to reproduce this work can be found in: \url{https://github.com/rraikman/popsynth-dotproducts}. This study made use of the BBH formation channel models released by \citet{2021ApJ...910..152Z} on \url{https://zenodo.org/record/4947741}.

\section*{Acknowledgements}

The authors thank Christopher Berry and Michael Zevin for their valuable comments and suggestions on the manuscript. SSB and TF were supported by the Swiss National Science Foundation grants PP00P2\_211006 and CRSII5\_213497.

\bibliography{sample631}{}
\bibliographystyle{aasjournal}

\end{document}